# *Biomechanics of Kuzushi-Tsukuri and Interaction in Competition*

*(A new global didactic Judo vision)*

*Pictures Courtesy by David Finch*

By Attilio Sacripanti [1,2,3,4,5]

1ENEA (National Agency for Environment Technological Innovation and Energy) Robotic Laboratory
2University of Rome II "Tor Vergata" Italy
3 FIJLKAM Italian Judo Wrestling and Karate Federation
4 European Judo Knowledge Commissioner
5European Judo Didactic Commission Scientific Consultant

**Abstract**

In this paper it is performed the comparative biomechanical analysis of the Kuzushi (Unbalance) – Tsukuri (the entry and proper fitting of Tori's body into the position taken just before throwing) phases of Judo Throwing techniques.

The whole effective movement is without separation, as already stated by old Japanese biomechanical studies (1972 -1978), only one skilled connected action, but the biomechanical analysis is able to separate the whole in didactic steps called Action Invariants.

The first important finding singled out is the existence of two classes of Action Invariants the first the General one' connected to the whole body motion is specific of shortening distance in the Kuzushi Tsukuri Phase.

The second one, the Specific Action Invariants is connected to the superior and inferior kinetic chains motion and right positioning connected both to Kuzushi and Tsukuri phases.

Some interesting findings derive from this analysis: among throwing techniques, couple techniques could be independent from Kuzushi; instead physical lever groups need all two action invariants to be performed successfully and as complex motor skill they are more difficult to perform than the first ones.

Complexity in motion is connected to the energy consumption, and to the skill ability of athletes.

This biomechanical comparative analysis is useful from the didactic point of view, clarifying the basic mechanical steps of all throws, and showing also hidden connections, similarities and differences among throwing Judo techniques.

The last part of the paper concerns the study on Interaction in Competition, performed by the author, showing 20 years results on Judo Biomechanics.

1. ***Introduction (Historical remarks)***
2. ***Biomechanical Improvement and Action Invariants***
3. ***General Action Invariants (GAI)***
4. ***Specific Action Invariants (SAI)***
5. ***Throws example emanating from General and Specific Action Invariants***
6. ***Throws example emanating from General Action Invariants only***
7. ***Biomechanics of Tsukuri Movements a mathematical remarks***
8. ***Interaction in Competition***
9. ***Conclusions***
10. ***Bibliography***

# *Biomechanics of Kuzushi-Tsukuri and Interaction in Competition*

*(A new global didactic Judo vision)*

*Pictures Courtesy by David Finch*

By Attilio Sacripanti[12345]
1ENEA (National Agency for Environment Technological Innovation and Energy)
2University of Rome II "Tor Vergata" Italy
3 FIJLKAM Italian Judo Wrestling and Karate Federation
4European Judo Knowledge Commissioner
5 European Judo Didactic Commission Scientific Consultant

## *1. Introduction (Historical remarks)*

Biomechanics is powerful tool able to group many body's actions under very few principles, in such optics we can consider the Dr. Kano approach to the classification of techniques really as a proto-biomechanical work.

Many different movements able to throw grouped under four classes (Te waza, Koshi waza, Ashi waza, Sutemi waza) namely ( hand techniques, hip techniques, leg/foot techniques, techniques by sacrificing personal balance to throw).

Another similar consideration could be applied to the Kuzushi – Tsukuri - Kake division; (Unbalance, Entry and proper fitting of body into the position taken just before throwing, Throwing action) in scientific way it is the differential method applied to one complex movement for teaching reasons, the method of division of complex patterns in more simple interconnected steps is actually wide utilized as tool in modern biomechanics of sports.

Then at this point it is possible to ask the question: could Dr. Kano know the biomechanics?

I don't know the answer, but modern Biomechanics was born in 1680 with the book of the Neapolitan scientist Alfonso Borelli "De Motu Animalium" and its methods spread around and we can find between 1850 and 1920 a lot of works applied to human movements, work movement and gymnastics in France, England, Germany, Soviet Union, United States and in its studies it would be possible that Kano meets some of these knowledge.

However we must remember that science is application of clear thinking or as F. S. Marvin affirms **"The essence of science is to discover identity in difference."** Then from that point of view Kano and co-workers applied a really scientific method to judo classification.

Judo competitions are from the Biomechanical point of view, an intriguing complex nonlinear system.

Around 20 years ago seeing the basic Matsumoto work on competition about the All Japan judo championships and giving a look at athletes shifting paths, and at their summation; I developed the idea of the Brownian motion of the couple of athletes system, but applying the differential analysis I tried to find also more simple approach to study complex competitions like judo sports.

Then the way that I found to study judo competition on the example of Dr Kano was to analyze competition divided in two steps: motion + interaction.

When you study MOTION you must analyze the system as a whole and it is possible to study the tracking paths of the Centre Of Mass (COM) on the Tatami that is a 2D trajectory projection of a 3D Brownian motion in space.

Then analyzing the motion of COM projection on the mat ( that is the only experimental data obtainable) I found a random walk continuous in time or better a Fractional Brownian Motion that is auto similar also, and from that was born my equation based on friction and Gaussian noise (push/pull forces in every direction).
Speaking about rotational approach, momentum of inertia and conservation of angular momentum we necessarily must jump in the couple system because these parameters are essential in INTERACTION.
Interaction is a macroscopic system and techniques, applied in it, are grounded both on a couple of forces and a physical lever, as basic tools to throw. In interaction for rotational kuzushi, Tai sabaki hando no kuzushi, etc. it is useful to apply the well known classical Lagrangian mechanics.
And among other results it is demonstrable that Uke falling trajectories are geodetics of some specific symmetry of the couple system; this means the less energy expenditure trajectories, because athletes unconsciously try to spend less energy in throwing.
During competitions athletes face against special situation continuously changing in time, these special situations, that are grounded on the infinite Athletes' relative body positions changing in time, must be studied and repeated during the specific post fight technical training.
These repetitions should be utilized by the athletes for learning the better way of governing such situations (INTERACTION).
This is the most important part of competition, in fact the analysis of managing the relative body's position that very often occur during the fights is essential to throw.

### 2. *Biomechanical Improvement and Action Invariants*

With the slow motion of the fight can be understood the state of athlete's technical preparation and to take data for his technical improvement. This aspect of Judo Match Analysis is worldwide part of the today technical training and improvement. The biomechanical problem of the technical improvement in Judo it is not only a problem of the athlete technical capability, but more often of the specific positional situation produced by the adversary in the couple system. From the biomechanical point of view it is more easy to consider the couple of athletes as a whole system in stable equilibrium in which the transition phase Kuzushi Tsukuri, depends strongly by the adversary position and action.
It is interesting to remember at this moment the nine classes singled out in Japanese judo to define the right Tori action utilized to overcome the defensive standing of Uke.

1. ***Tobi Komi (jumping in )***
2. ***Mawarikomi (spinning in)***
3. ***Hikidashi (pulling out)***
4. ***Oikomi (dashing in)***
5. ***Daki (to hug holding)***
6. ***Debana (Thwarting the opponent)***
7. ***Nidan Biki ( two stage pull)***
8. ***Ashimoki (leg grab)***
9. ***Sutemi ( body drop).***

This qualitative nomenclature, connected to some specific motor actions, is suitable to apply the Kuzushi -Tsukuri phase in real competitive positioning overcoming the arms resistance grips). But today with the growing of scientific studies in the world, this old approach, is ousted from very update analysis utilizing advanced technologies like "Biogesta saga 3D or Vicon system" to study not only kinetics and kinematics of techniques, but also to single out the “***Action Invariants***” relative to each technique to apply very effective Kuzushi Tsukuri phase.
In the following pictures there are shown two powerful Kuzushi-Tsukuri actions finishing into a Kake phase the first: couple of forces, and the second ones:  physical lever.

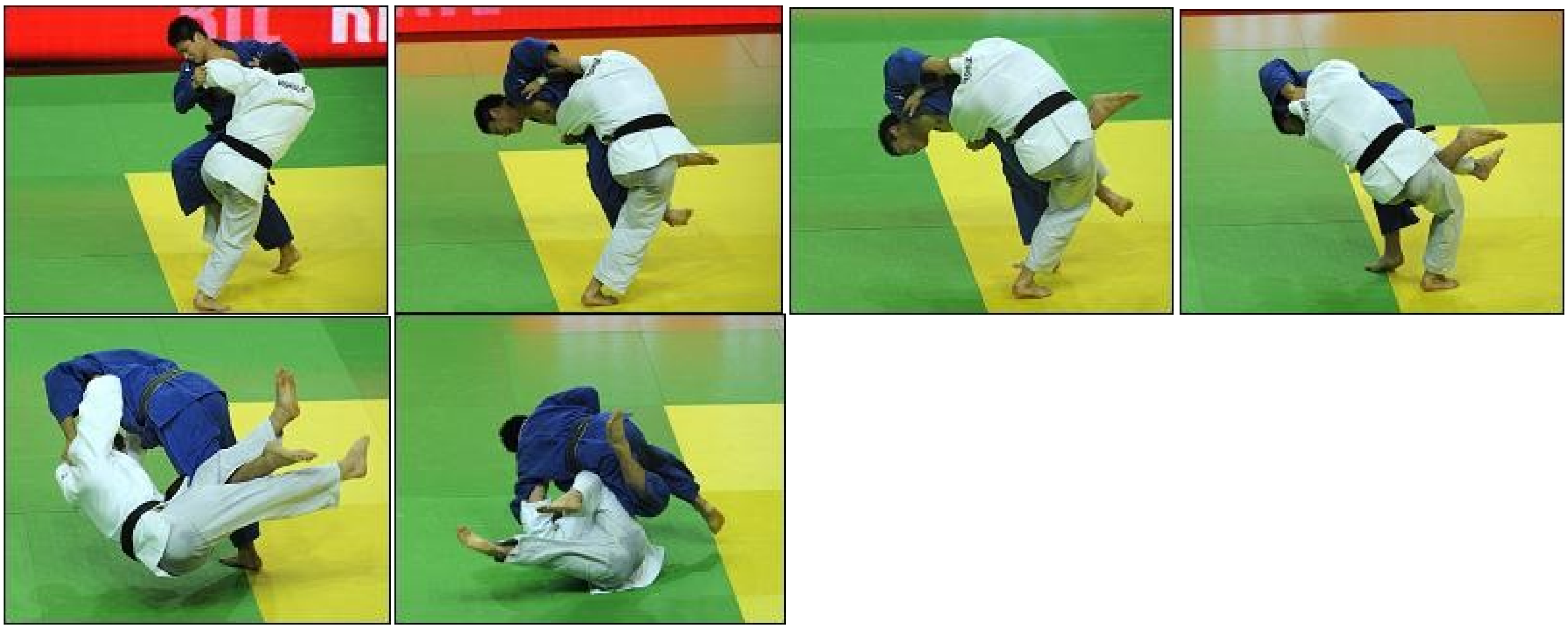

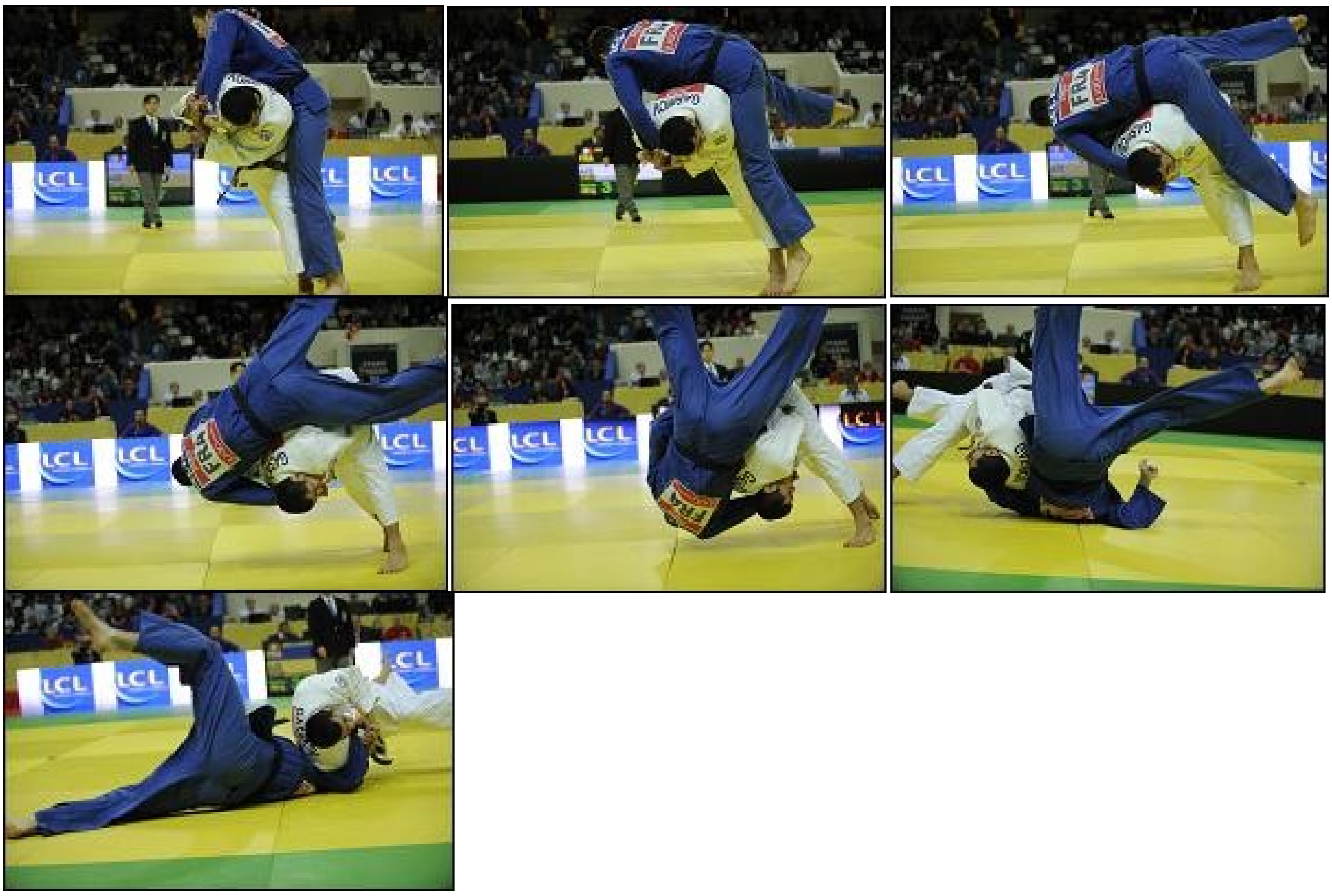

***Figg.1&2 Kuzushi-Tsukuri- phases flowing into couple of forces and physical lever Kake phases.***

In the following there are shown very interesting results obtained by a French Study on Ippon Seoi Nage, similar results could be extrapolated from every Judo throw.

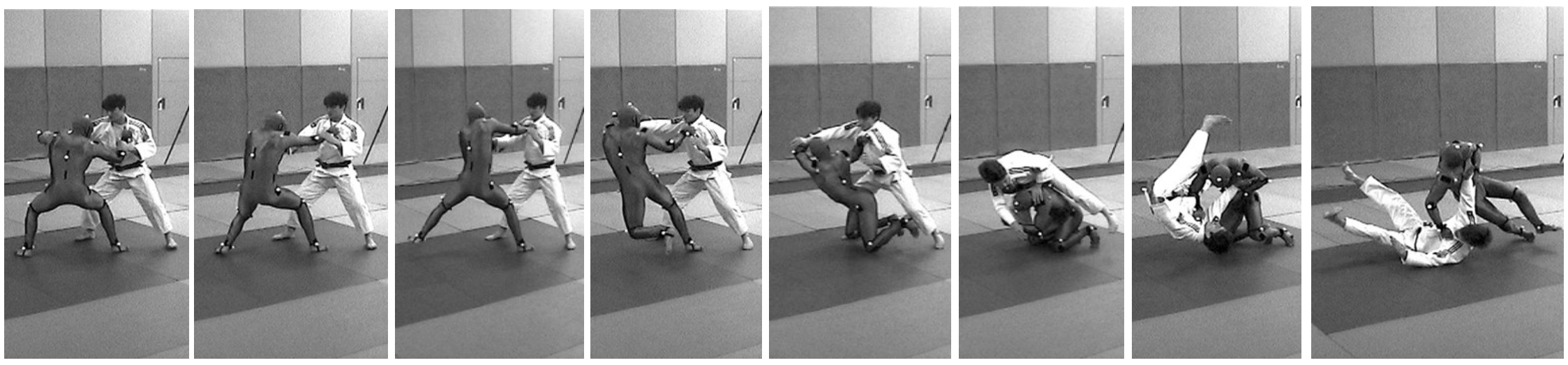

*Fig.3 Advanced biomechanical researches on Suwari Seoi by Saga 3D (Poitiers University Fr.)*

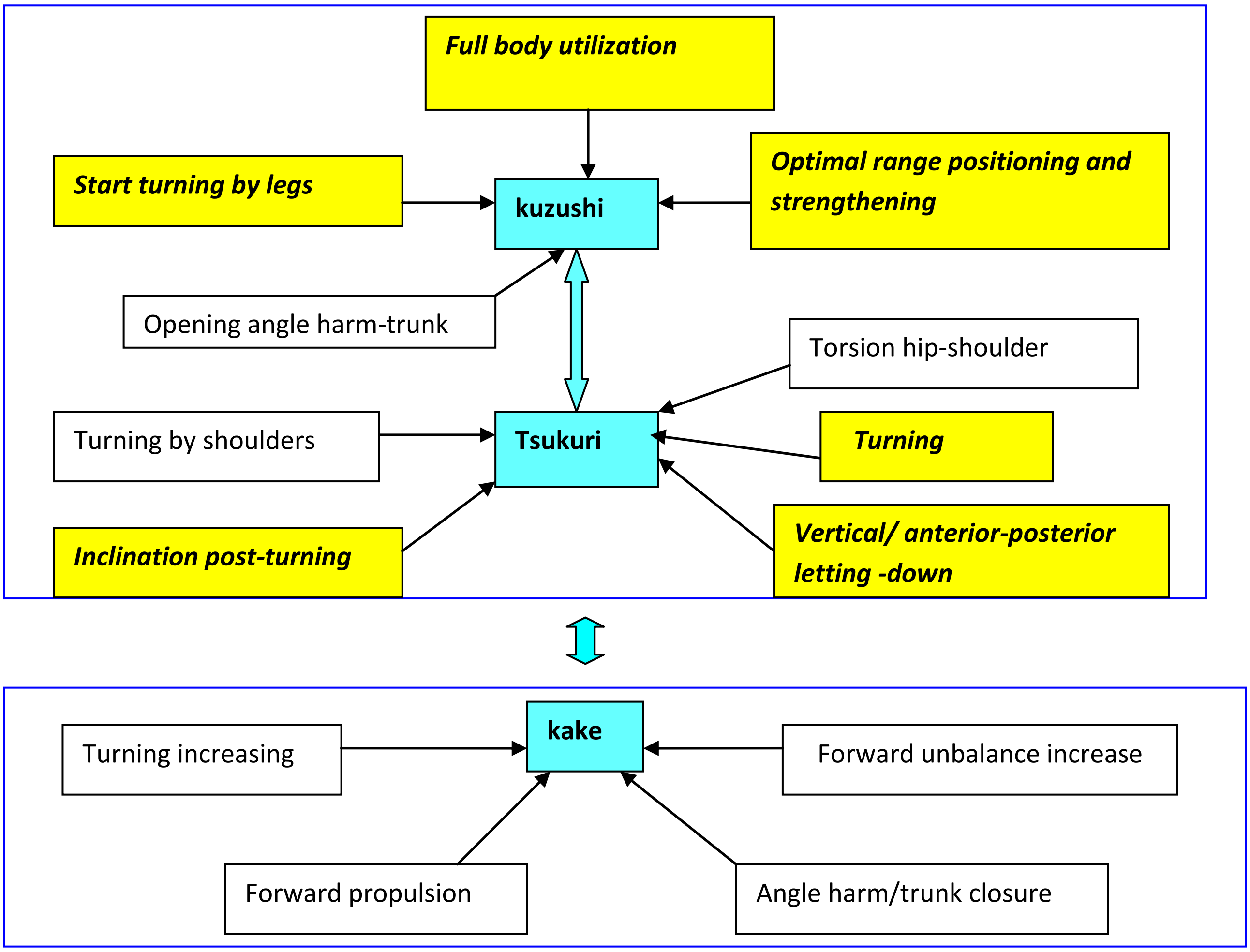

***Diag.1, Suwari Seoi "Action Invariants" identified by French analysis***

from: ***Blaise & Trilles Comparative mechanical analysis of the same judo throwing : Seoi Nage, realized by five experts of the Judo French Federation - Science & Motricité n° 51 —49-68 -2004/1***

The whole Kuzushi Tsukuri Kake effective movement is without separation, as already stated by old Japanese biomechanical studies (1972 -1978), only one skilled connected action, but by the biomechanics it is possible to separate the whole in didactic steps called ***Action Invariants.***

**Action invariants** are very similar movements which are possible to find every time in the Throws' Kuzushi-Tsukuri phase.

It is interesting that in biomechanics these **Actions Invariants** should be traced back to the Hamilton –Lagrange Equation and to the Hamilton Action principle.

$$S(q,t)=\int L(q,\dot{q}$$

With S = the Action (throwing techniques); and L= the Lagrangian (essentially the energy) of the system. In fact if we consider in a first approximation constant the external energy of the system (gravitational field) it is possible to write:

$$S(q,t)=W(q)-Et=L(q,\dot{q}$$

$$\delta S(q,t)=\delta\int_{t_1}^{t_2} L(q,\dot{q} \qquad \int_{t_1}^{t_2}\left[\frac{\partial L}{\partial q}-\frac{d}{dt}\left(\frac{\partial L}{\partial \dot{q}}\right)\right]\delta q\,dt=0\Rightarrow$$

$$\frac{\partial L}{\partial q}-\frac{d}{dt}\left(\frac{\partial L}{\partial \dot{q}}\right)=0$$

*if the system is not conservative, we must write*:

$$\frac{\partial E}{\partial q}-\frac{d}{dt}\left(\frac{\partial E}{\partial \dot{q}}\right)+Q=0$$

According to Hamilton's principle, the true evolution of S(q,t) is an evolution for which the action is stationary (a minimum, maximum, or a saddle point).

Normally in Judo Action it is asked for the minimum; then this is the called the principle of the minimum action. In Judo terms the ***Action Invariants***, should be recognized as the minimum path, in time, of body's shift to acquire the best Kuzushi- Tsukuri position for every Judo Throws.

Generally this is possible in a conservative field, if we consider a non conservative field, it is necessary to consider for the global balance the heat Q emitted, and in this situation it is not possible to find a minimum of the action.

But in judo competition such movements are so fast ( 0.6 sec see tab 3) that interaction ( complete throws) in the couple of athletes system could be judged, with very good approximation, adiabatic (without thermal variation during this fastest action) then it is possible to apply the principle of minimum action to every fast throw in competition and from that the name of **Action Invariants**.

In the case where it is possible to find a minimum, the two following biomechanical principles are true:

a) ***Best is the Judo Technique, minimum is the Athletes' energy consumption.***

b) ***Best is the Judo Technique, minimum is the Athletes' motion path for positioning.***

The modern match analysis tends on deeper the study of the most important competitive aspect of throwing: the Kuzushi – Tsukuri phase.

A lot of scientific studies have been developed in the world to understand the best way to achieve the better relative position inside the couple of athletes systems; from the older to the modern ones' there are to remember, for example:

Asami T., Matsumoto Y, Kawamura T., **Studies on Judo Techniques with special reference to Kamae and Kuzushi** , Kodokan report ,IV, 1972. Matsumoto Y, ,Takeuchi Y, Nakamura R., Tezuka M., Takahashi K., **Analysis of the Kuzushi in Nage Waza** Bulletin of the Kodokan report ,V, 1978.

**Analysis of Different Tsurite Movements of Elite Judo Competitors** in the 2005 bulletin for the association for the scientific study on judo of the Kodokan; from Akitoshi Sogabe (Konan

University) and others, **The Biomechanics of Loss of Balance in Olympic Sport Judo, possibilities of measurement of biomechanical parameters** by Nowoiski of the Olympic Centre in Hanover Germany 2005, or **A biomechanical Investigation of kuzushi of O soto gari in Kano Cup international competition** from Komata,and co-workers 2005 or
**Kuzushi and Tsukuri and the Theory of Reaction Resistance** from Rodney .Imamura (California State University Sacramento) and Iteya in the Kodokan bulletin 2007. **Biomechanical Analysis for Application of judo Techniques** Kim & Yung Korean J of Biomechanics 2000 .
All these study are very interesting and deep in their evaluation, but all these quantitative results are very often connected to the specific technique or movements and it is difficult to generalize such findings to a global Judo understanding.
Based, sometime, on the few numbers of samples, it is very difficult and hard to generalize at global level, deep and interesting data like these very beautiful results obtained in some of the previous studies.

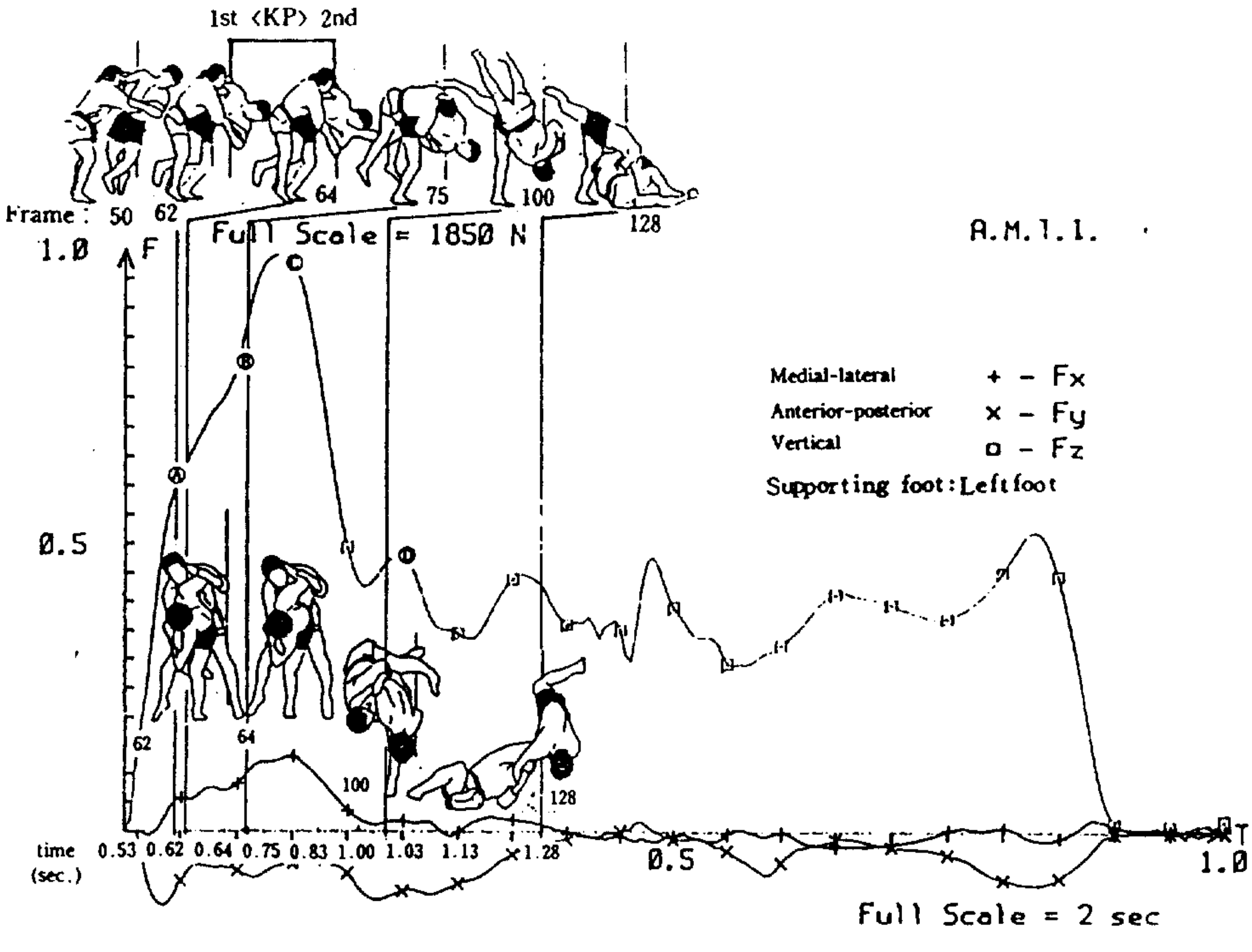

***Fig 4 Ground reaction forces in space for Uchi Mata (from Korea)***

From *Biomechanical Analysis for Application of judo Techniques Kim & Jxjing 2000.*

| Items | Groups | Sample # | Mean | S.D. | t-value |
|---|---|---|---|---|---|
| Leg angle | Knee-extended | 18 | 55.13 | 6.63 | |
| | Knee-flexed | 18 | 88.60 | 18.36 | -7.27* |
| Body angle | Knee-extended | 18 | 94.95 | 5.81 | |
| | Knee-flexed | 18 | 89.86 | 2.69 | 3.36* |
| Tsukuri time | Knee-extended | 18 | 0.214 | 0.032 | |
| | Knee-flexed | 18 | 0.202 | 0.031 | 1.12 |
| Raise leg time | Knee-extended | 18 | 0.243 | 0.041 | |
| | Knee-flexed | 18 | 0.240 | 0.019 | 0.261 |
| Sweeping time | Knee-extended | 18 | 0.158 | 0.016 | |
| | Knee-flexed | 18 | 0.170 | 0.017 | -2.179* |
| Kake time | Knee-extended | 18 | 0.401 | 0.037 | |
| | Knee-flexed | 18 | 0.410 | 0.026 | -0.816 |
| Movement time | Knee-extended | 18 | 0.616 | 0.054 | |
| | Knee-flexed | 18 | 0.613 | 0.038 | 0.178 |
| Force of hands | Knee-extended | 18 | 0.838 | 0.144 | |
| | Knee-flexed | 18 | 0.847 | 0.230 | -0.131 |
| Force of sweeping | Knee-extended | 18 | 0.603 | 0.147 | |
| | Knee-flexed | 18 | 0.696 | 0.099 | -2.235* |

**Table 1 Mean, Standard Deviations, and t-test, Results for Movement Time and Force of Knee-extended O Soto Gari and Knee-flexed O Soto Gari* p _ 0.05.**

Kwei-Bin Kuo *Comparison between knee-flexed and knee-extended styles in Major outer leg sweep* Biomechanics Symposia 2001 University of S. Francisco

Separation distance

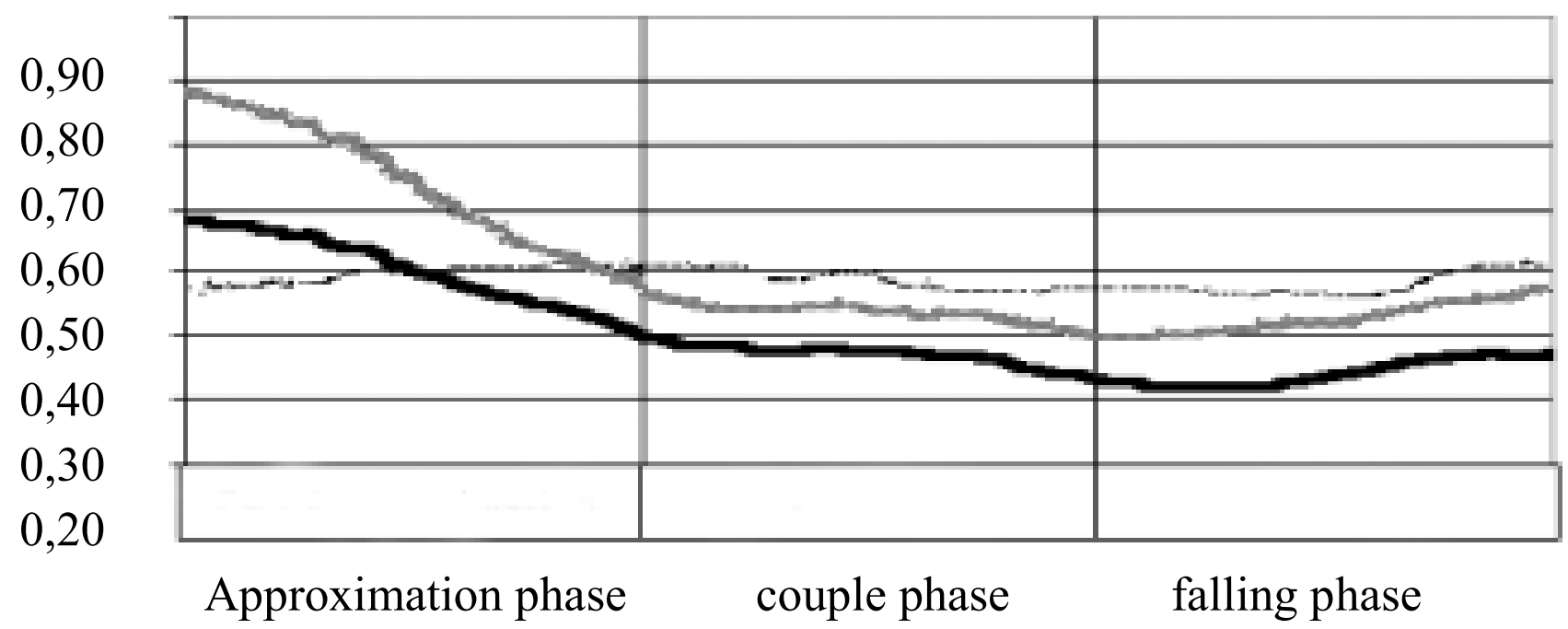

Diagram ***2 separation distance in Uchi Mata (from Spain)***

Suarez, G.R., Davila, M.G. *Análisis de los factores que determinan la inestabilidad del sistema tori-uke en la técnica del uchi-mata en judo*. Biomecánica, 10 (1), pp. 25-32, 2002.

From the biomechanical point of view it easy to understand that the problem to shorten the relative distance between the Athletes into the couple system it is essential to perform a very effective Kake phase. Remembering that the biomechanical classification of throws in judo is able to group all techniques (considering only the Kake phase), under two specific physical principles. (See Fig.1 and 2) This is a clear example of powerful capability of biomechanics to simplify and group in classes different complex sports movements (from four to two). Obviously the Kuzushi-Tsukuri phases are blended and connected in a whole movement, but for teaching it is better to separate them in two steps Kuzushi and Tsukuri. Now applying the biomechanical analysis to the Tsukuri phase, it is possible to group these almost infinite complex movements able both to shorten the distance between athletes, and to acquire the right body's posture to throw, in few general classes.
One of the simplest ways to classify Tsukuri movements is based on the singling out some general simple body movements called ***General Action Invariants***: these movements related to the distance shortening between athletes that could be analyzed as Tori's body movements.
Now it is interesting that appears the real difference between the goals of Kuzushi- Tsukuri, in fact the Kake phase of couple of force doesn't need any refinement movement; instead the physical lever phases needs the addition of some specific movements of refinements depending by the right positioning action of both kinetic chains. This is the main difference between techniques that flow into Kake by Couple or Lever. All the techniques applying the physical lever must be refined by means of so called ***Specific Action Invariants*** that are common to each class of these throws related both to the effective positioning of kinetics chains and the way to apply forces to unbalance Uke before throwing. In such way the classes ***General Action Invariants*** are connected to both physical principles that are basic for one effective Kake phase, but the sparkling difference is that they are connected Directly to the Couple of forces and Secondarily connected by means of ***Specific Action Invariants*** to the techniques of the physical lever Kake phase.
And this special finding have a strong new information for us, that will be explained in the two next paragraphs.

### *3. General Action Invariants (GAI)*

Under this name, we collect the whole body movements useful both to shorten the distance and to place in right way the body relatively to adversary's body position.
The biomechanical analysis of this class of invariant, starting from one of the six ***Competition Invariants***,( see *Advances in Judo biomechanics research*) specifies that there are only three ***General Invariants*** in Kuzushi-Tsukuri phase:

1st Distance shortening without rotation.
2nd Distance shortening by a complete rotation clockwise/counter-clockwise (0°to180°).*
3rd Distance shortening by half rotation clockwise/counter-clockwise (0° to 90°)*

To all these ***General Invariants*** are connected both couple and lever in Kake phase, the choice of the Kake physical principle determines only for one principle the connection with the ***Specific Action Invariants*** that follow the general ones'.

**These angles are valid for still standing athletes ( study)but in competition they could change if the athletes are moving, a classical example is the Japanese well Known Hando No Kuzushi ( see fig2 ).*

### 4. *Specific Action Invariants (SAI)*

Then if we consider only the Techniques with a Kake phase of physical lever, soon connected to the ***General*** ones' are, before the Kake phase, the ***Specific Action Invariants*** that in general in the Japanese lexicon, are responsible for the different names of throwing techniques.
In these techniques must be present all two phases Tsukuri-Kuzushi
The ***Specific Action Invariants applied*** by kinetic chains must be divided in two sub classes :

a) ***Superior chain Action Invariants (SSAI)*** (generally this class is responsible of forces application)
b) ***Inferior chain Action Invariants ( ISAI)*** (generally this class is responsible of fulcrum application)

The ***SAI*** are infinitive then, the only way to classify them in collective groups, is based not in defining angles or direction but in singling out the specific goal.
In general considering the degrees of freedom connected to the kinetic chains, related to body movement, we can easily understand that all the potential movement in the ***Specific Action Invariants,*** for the superior chain, in the Kuzushi actions are connected to the three degree of freedoms of the Achromium Joint (shoulder).
From the other side the most important part of the inferior actions are in charge of two Joints hip and knee, less part to the system foot/ankle ( the first for setting better his ones' body in relation to Uke' s body, and the second ones to applying only fulcrum in the lever ).
Another important notation is that both ***Specific Action Invariants***, must be connected each other in one harmonic way to produce the right and effective throw result.
Kake phase, for the techniques of physical lever, is the result of the well interconnected work performed by both kinetic chains in different time steps: at first the superior chain open space for the body in the adversary grips, then the ***General Action*** (distance shortening) is developed followed harmonically by the connected work of both ***Inferior*** and ***Superior Action Invariants*** through the abdominals and trunk muscles.
These techniques need more skill in harmonic chains connected movements, than couple techniques; in fact often they are ineffective because lack in harmony into one of the previous movements is able to stop the throwing result.

### 5. *Throws example emanating from General and Specific Action Invariants*

Then resuming, it is possible to have about denomination, for example:
The same ***General Action Invariant*** ( **full rotation**) that ( considering only the inferior chains) splits by three ***Inferior chain action invariants*** from up to down, into three well known techniques Seoi nage standing , Seoi Otoshi and Suwari Seoi.
All these Kuzushi-Tsukuri movements flow in the Kake phase of physical lever application with variable arm, from the most energy wasting one to the lesser ones.( See fig. 5,6)

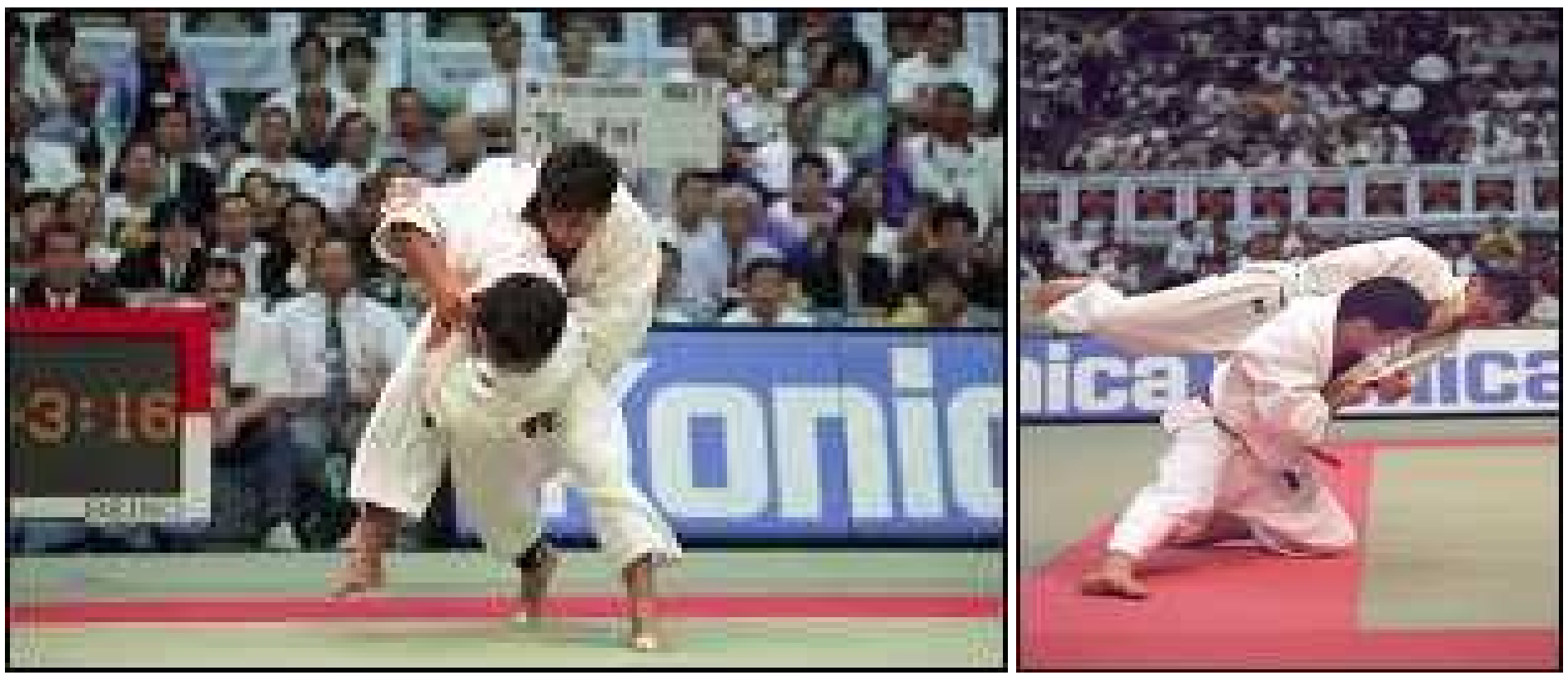

***Fig 5,6  Inferior Chain Action Invariant (Seoi); Inferior Chain Action Invariant (Seoi Otoshi);***

The same ***General Action Invariant*** (**direct shortening without rotation)**, that ( considering only the inferior chains) splits by three ***Inferior chain action invariants*** from down to up, into three well known techniques , Sasae Tsurikomi Ashi, Hiza Guruma, and Tomoe Nage.
 If you look at the inferior kinetic chain only leg and foot of Tori go up from the ankle, to knee, to Uke's abdomen.  All these Tsukuri movements flow in the Kake phase of physical lever application with variable arm, from the less energy wasting one to the most expensive ones.
The same ***General Action Invariant*** (**full rotation**) that (considering only the inferior chains) from down to up,  splits by three ***Inferior chain action invariants*** into three well known classic techniques Tai Otoshi, Ashi Guruma and O Guruma.
If you look at the inferior kinetic chain only leg and foot of Tori go up from the ankle, to knee, to Uke's abdomen.  All these Tsukuri movements flow in the  Kake phase of physical lever application with variable arm, from the less energy wasting one to the most expensive ones.
In the next figures there are shown some inferior chain action invariants connected to Sode Tsurikomi Goshi

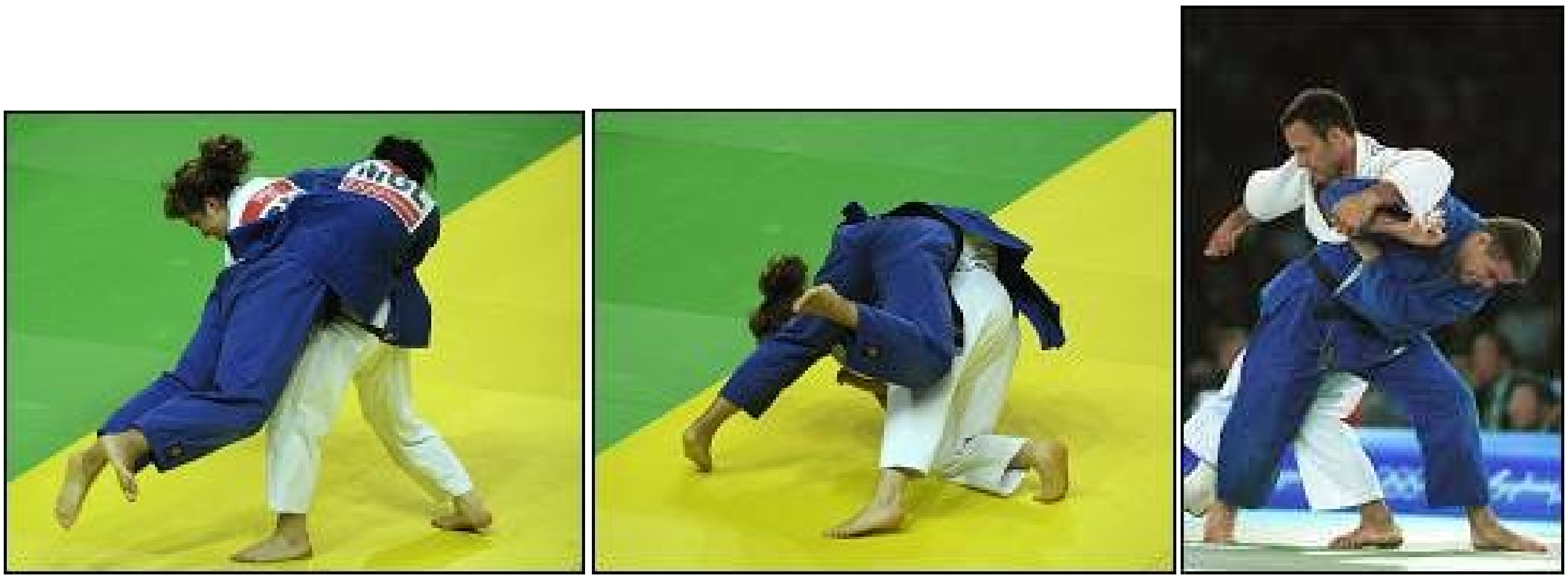

***Fig 7,8,9  Inferior Chain Action Invariant for pseudo Sode Tsurikomigoshi;***

 Among the lever group techniques there is a Maximum harm subgroup in which the ***Superior chain Action Invariants*** play the most important role there are techniques like: Uki Otoshi or Sumi Otoshi in which there is no body contact at all, but the fulcrum, under Uke feet,  is made by the friction between mat and foot.

In the next figures ( Fig 10) we can see the main use of the ***Superior chain Action Invariants*** (generally this class is responsible of forces application) and in this way it is possible to define the main useful angles for Kuzushi that is essential in these kind of techniques, because as it is easy to understand it is not possible to apply a physical lever principle, without the application of force far from the fulcrum. Not only, but the presence of the connected movements makes this techniques biomechanically speaking more complex, and on the basis of the physical principle for their useful application in competition whatever shifting velocity the couple of athletes has, they need of a stopping time ( also very short) to be applied.

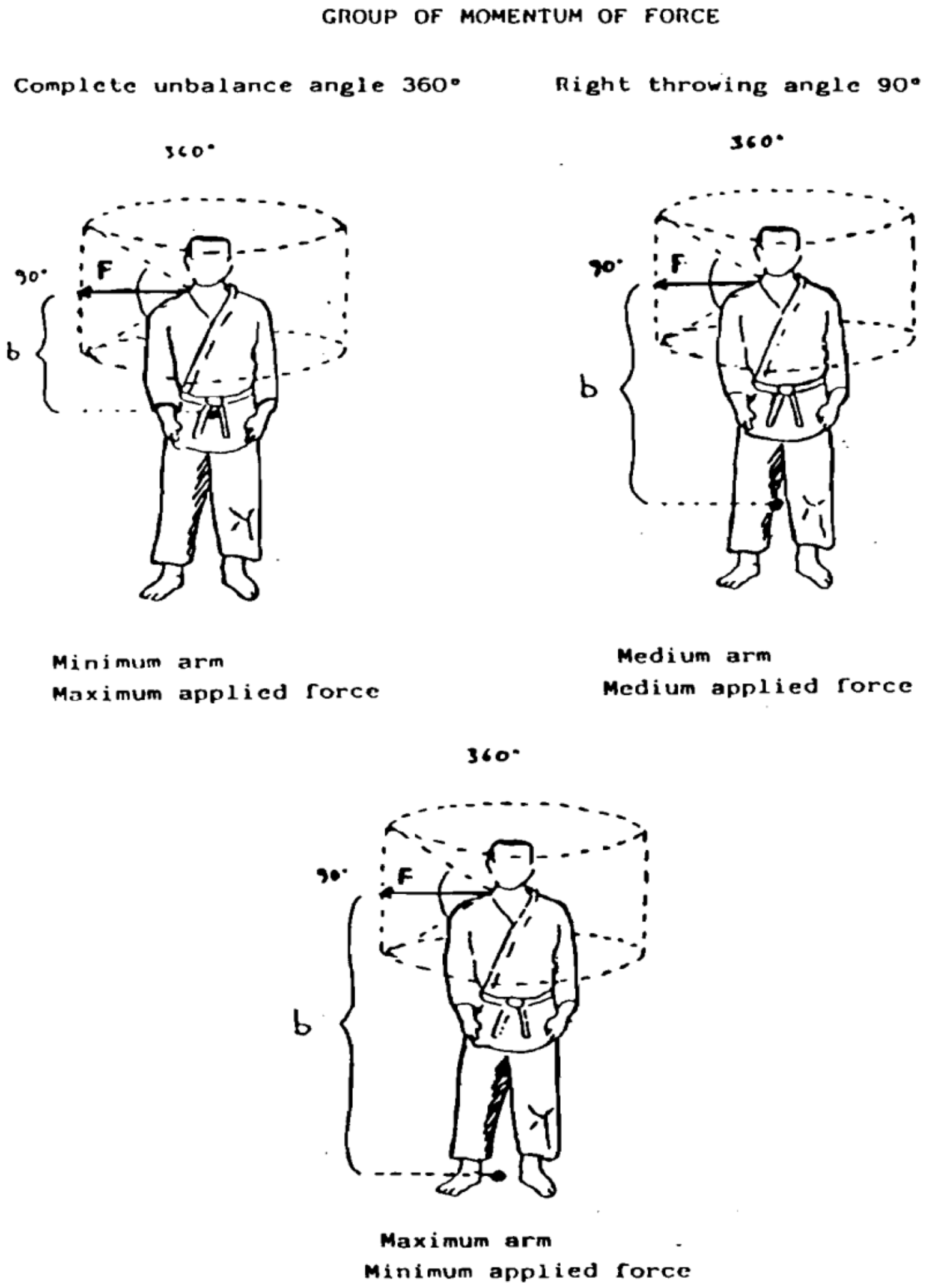

***Fig 10  Complete and right angles to unbalance and throw Uke with physical lever.***

Sacripanti A, *Advances in Judo Biomechanics research* VDM Verlag 2010 *ISBN* -10: 3639105478 -*ISBN*-13: 978-3639105476

### 6. *Throws example emanating from General Action Invariants only*

About the ***General Action Invariant*** as seeing before they are directly connected to Tsukuri Kake Phase and the fantastic information is that in this case Kuzushi is not a necessary and sufficient condition, and it could be non present in some couple techniques application. ( see Figg. 11,12)

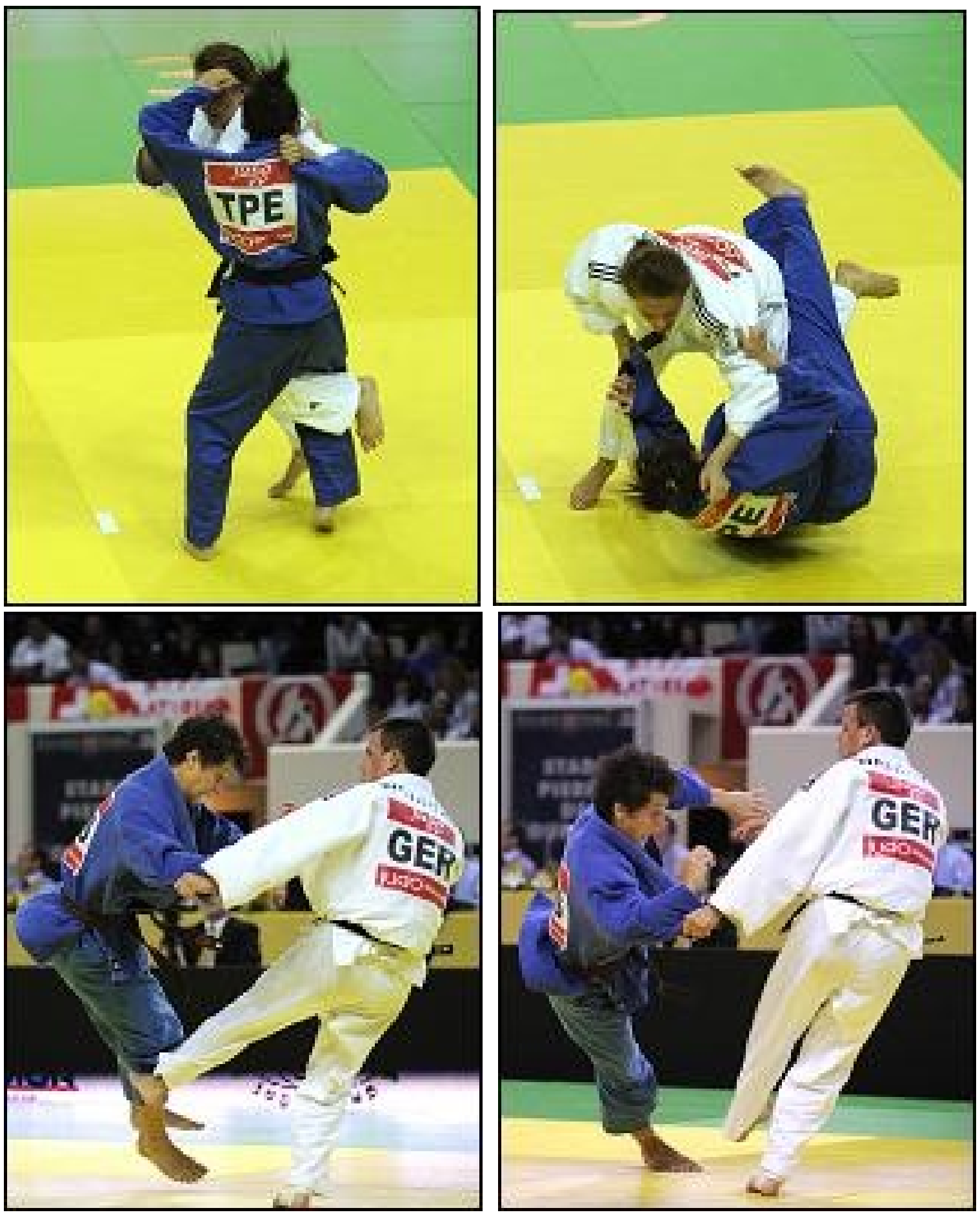

***Fig 11,12 General Action Invariant flowing in Kake phase without Kuzushi (O Uchi Gari - Okuri Ashi Arai);***

This means that in such group there are necessarily only Tsukuri-Kake phases, in fact the presence of Kuzushi helps obviously the throws but it is not absolutely necessary, because for Uke in unstable equilibrium, the rotation around his COM is helped by gravity force. The other valid information singled out is that, these more simple techniques, biomechanically speaking, can be applied in competition whatever shifting velocity the couple of athletes could have.
The couple techniques are connected to their Japanese denomination considering the Tori three body's symmetry planes in which lie the couple of forces. considering the first of the previous three we can have:

**Sagittal Symmetry Plane** Application of
$1^{st}$ ***General Action Invariants***: O Soto Gari, reverse direction Mae Ushiro Uchi Mata,
$2^{nd}$ ***General Action Invariants:*** Harai Goshi inverse application Ushiro Hiza Ura Nage,
$3^{nd}$ ***General Action Invariants:*** Uchi Mata, inverse (Back) Ushiro Uchi Mata

**Transverse Symmetry Plane** Application of:
3nd ***General Action Invariants:*** O Uchi Gari, Ko Uchi Gari (There are not inverse but opposite from the other side)
**Frontal Symmetry Plane** Application of:
1st ***General Action Invariants:*** Okuri Ashi Harai; opposite Okuri Ashi Harai from the left side,
3nd ***General Action Invariants:*** Ko Soto Gari

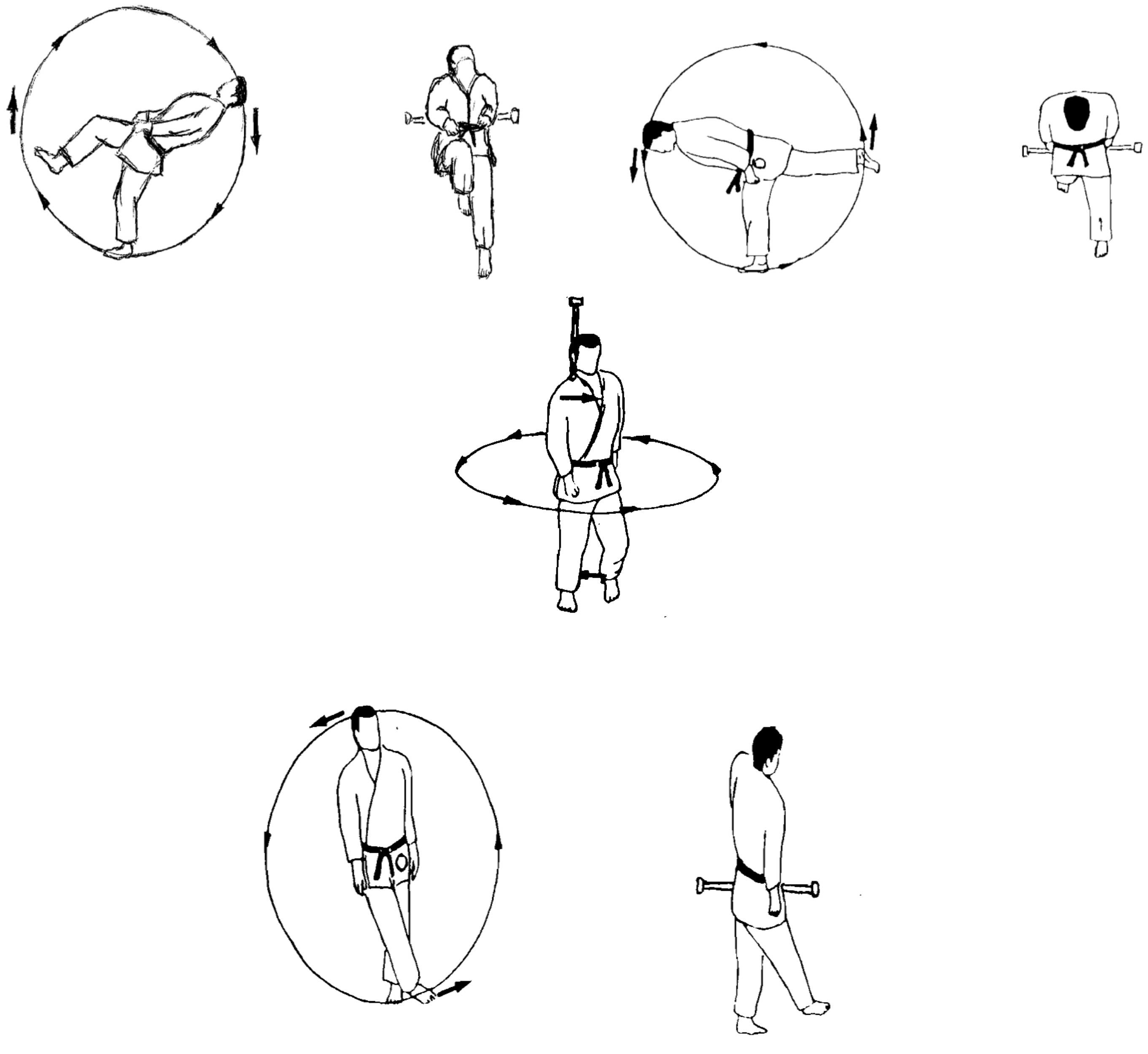

***Fig13-16 . Application of Couple in the three planes of Body symmetry: Basic Movements A) Sagittal, B) Transverse C) Frontal***

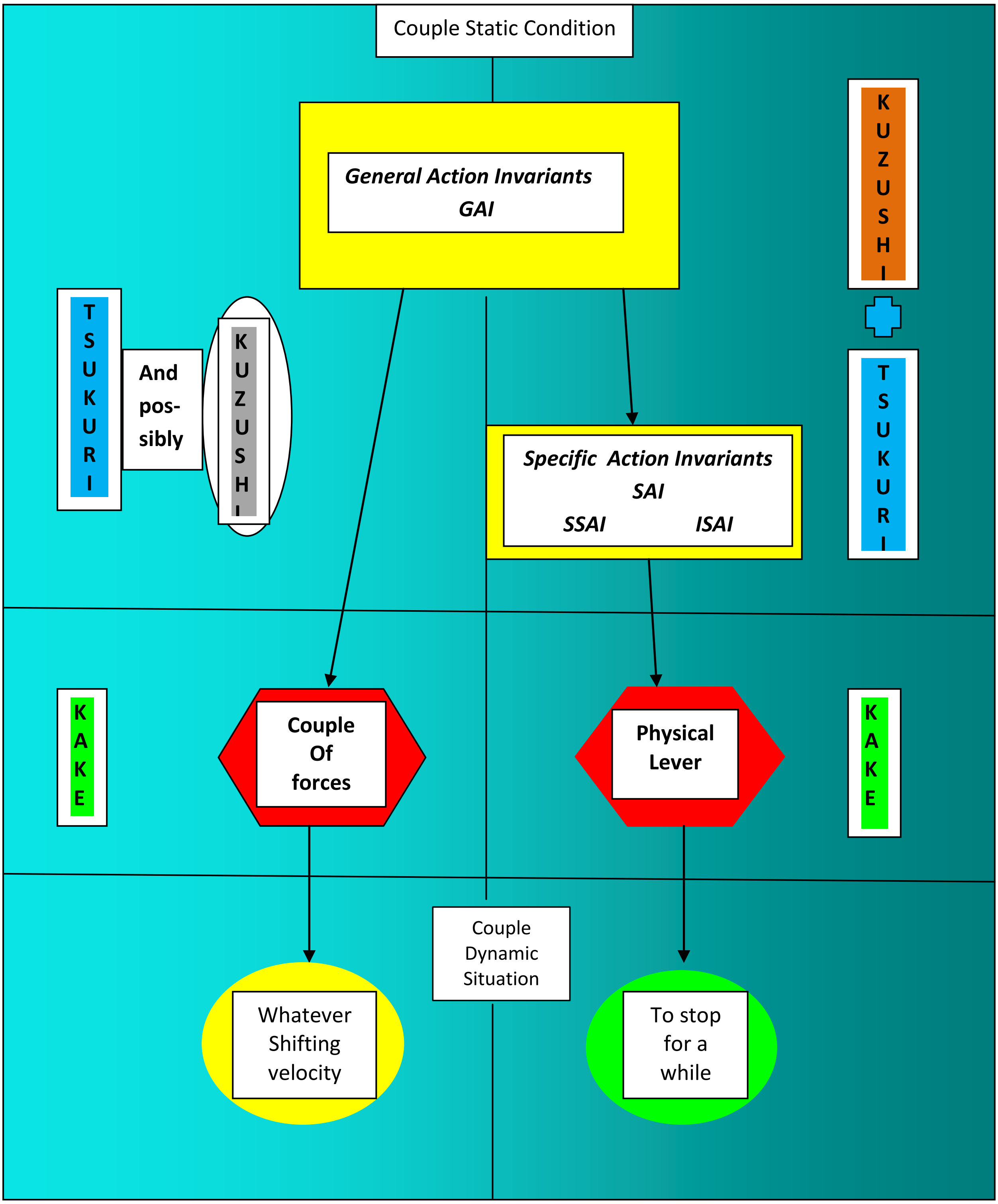

***Diag 3 Summary of the Kuzushi Tsukuri Action Invariants connected to Kake pahses***

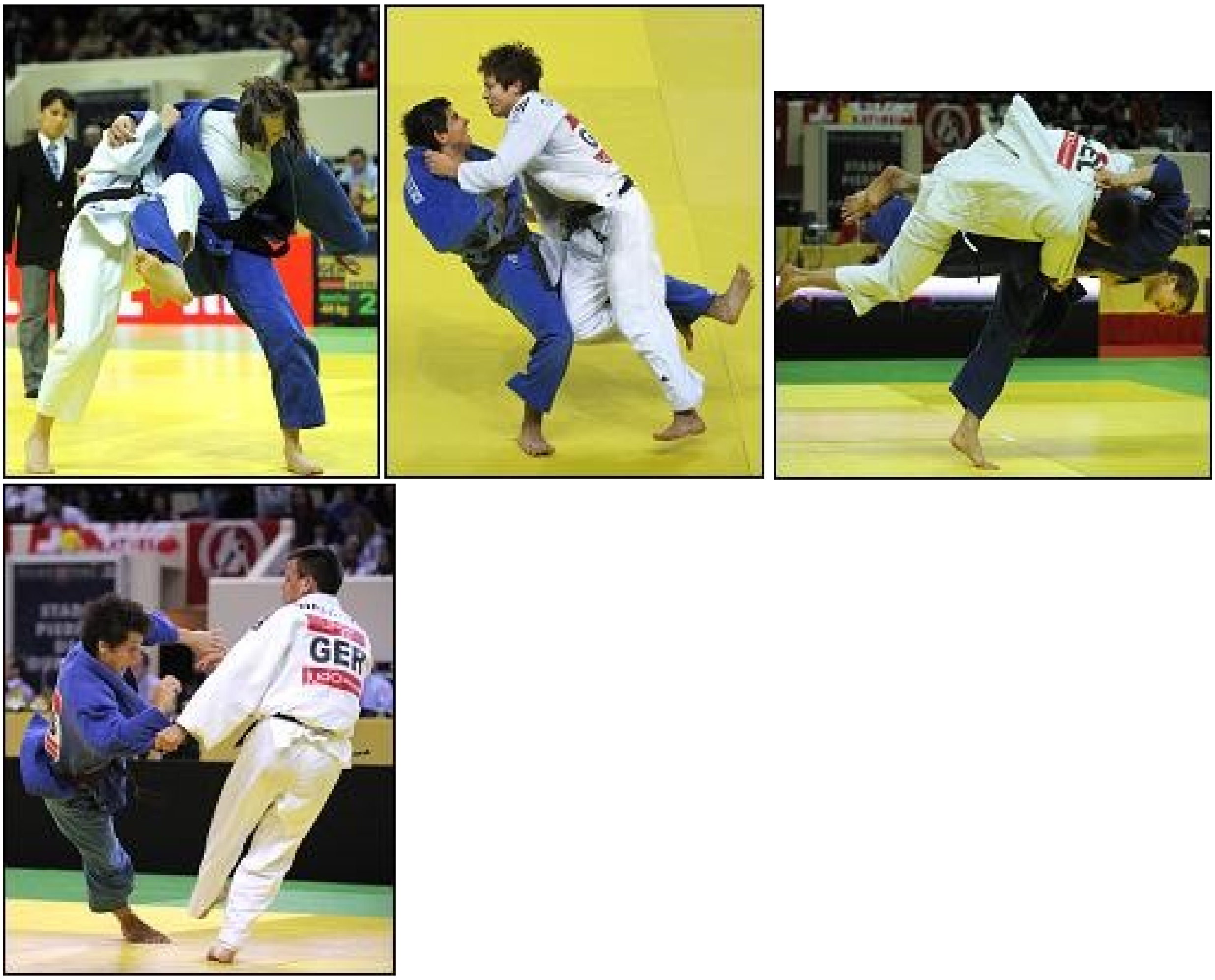

***Fig 17-20 Application of Tsukuri (Kuzushi) flowing in Couple of forces Kake phase into the three planes of body symmetry***

### 7. *Biomechanics of Tsukuri Movements a mathematical remarks*

The solution of Tsukuri movements lies in the field of Rigid Body Dynamics, the mathematical solution pass through the resolution of Euler equations.
Obviously the ***General class of action invariants*** is very simple movements, in training condition, but in real situation athletes must overcome the obstacle of the grips grounded on the strength and position of the arms and feet position ( ***Competition Invariants***) . The almost infinite situations arising from grips/feet relative positions are not explained in this paper, but are classed in the six classes of ***Competition Invariants***, and the mathematical and analytical solution of this problem is very hard and almost impossible to solve with the classical mechanical approach.
Generally speaking during competitions the problem to overcome the grips obstacle, is very often based on a well known Japanese trick, the techniques connection like Ko Waza into O Waza or inversely O waza into Ko Waza ( little techniques open space for big techniques, and inversely).
During the comparative analysis of competitions it is easy to see Japanese athletes apply Ko Waza (little technique) Couple Techniques like Ko Uchi Gari without Kuzushi to obtain a Kuzushi [Hando no Kuzushi] position useful to apply O Waza,( big technique) Lever Techniques, like Seoi Nage for example. Instead Western player more often they start with O Waza, Lever Techniques, with no so good Kuzushi, like standing Seoi Nage, to stop the adversary and apply Ko Waza , Couple Techniques, without Kuzushi like Ko Uchi Makikomi for example.
Many other example are possible in techniques connection like Japanese O Uchi Gake into Uchi Mata and so on, but they are out of our paper's goal.
Coming back to mathematical application, also the general solution of ***General Action Invariants*** considering the infinite situations arising during a competition is very far from an analytical solution, but the geometrical view makes some light on this very difficult problem and the solution could be accepted considering the still system connected to Athletes couple, taking in account that for the techniques of physical lever groups couple must stop for a while. And very often, with good approximation, the motion for the couple applying techniques lying in the couple of forces group could be considered uniform both linear and circular, in these cases with a change in the reference system we are able to bring back the system in still position.
In term of trajectories the first class of ***General Action Invariants*** are almost right lines with specific direction, normally the best right inclination is, in direction of both adversaries' sides, because human body structure is less skilled to resist in such direction. For example into the group of couple of forces, this means couple applied in the frontal plane: like sided O Soto Gari, or Okuri Ashi (harai –barai).
For the other two classes based on rotations, some interesting remarks come from the Poinsot geometrical description of a free forces motion of a body, in such case in fact the motion is like a rolling of the body inertial ellipsoid (without slipping) on a specific plane, remembering that the curve traced out by the point of contact on the inertial ellipsoid is called polhode, while the curve on the plane is called herpolhode.
In a case like judo player, the body is cylindrical symmetrical and the inertial ellipsoid becomes an ellipsoid of revolution, then the polhode is a circle around the Athlete symmetry axis, and the herpolhode on the plane (Tatami) is likewise a circle.
These results in the judo player reference system are real only in the case of free force motion, (it is true considering the couple motion as a whole) but in our case INTERACTION into the couple, as already remembered, there are the push/pull forces and the friction forces acting on.

The problem is almost not analytically solvable for his complexity. However the gross indication (likewise circular trajectories) available for our analysis by Poinsot description is quite acceptable as indication for real situations. In the technical group of couple of forces laying in the Sagittal and Frontal planes of symmetry the Kake phase is made effective by reaping or sweeping with Tori leg/foot applying part of the couple. Interesting is to study the contact mechanics between legs that are a soft collision with friction between two viscoelastic bodies (the Athletes' legs).
The problem is hard in mathematical formulation because we are in presence of collision of two non linear deformable (soft) bodies and it is possible to approach it by Lagrangian formulation of continuous mechanics of contact.
For example in this optics we can solve only the following equation of body's part motion:

$$\Delta_x P + \rho_0 F = \rho_0 \frac{\partial V}{\partial t}$$

Where in the Lagrangian or Body material reference P' is the Piola –Kirchhoff stress tensor of first type, ρ the mass density, F the forces in the leg, V the speed of the leg/foot.
Using the second Piola Kirchhoff tensor P''= $f$ S the previous equation of motion can be written as:

$$\Delta_x fS + \rho_0 F = \rho_0 \frac{\partial V}{\partial t}$$

In which we have $f$ the leg deformation gradient tensor, and S the constitutive material equation (that gives us information about the muscular elasticity of the leg).
The collision or contact is well described by Signorini problem that give us the contact relationship, the solution is grounded on the introduction of a changed variable in the Signorini problem , that the well known relationships $v_n \leq 0$ ; $f_{cn} \leq 0$ ; $u_n f_{cn}=0$ will be changed in:

$$A_1 \text{ (Tori's leg)} \begin{cases} d(x_1, x_2) \geq 0 \\ f_{cn1} \leq 0 \\ x_1 f_{cn1} = 0 \end{cases}$$

$$A_2 \text{ (Uke's leg)} \begin{cases} d(x_2, x_1) \geq 0 \\ f_{cn2} \leq 0 \\ x_2 f_{cn2} = 0 \end{cases}$$

Where $d(x_1, x_2)$ and $d(x_2, x_1)$ are the distance between $A_1(x_1)$ and $A_2(x_2)$ and the normal is $N_1=-N_2$ this problem is solvable in the approximation of an infinitesimal contact point.
When the surface of collision is extended, the problem solution must be evaluated for each couple of points $x_i$ - $x_j$ .
The complete contact law is a complex non smooth dissipative law including normally three statuses , no contact; contact with sticking; and contact with sliding.
The numerical solution to solve the equation of motion is grounded on the application of equation of motion in weak form.

$$\iiint (\Delta_x fS + \rho_0 F - \rho_0 \frac{\partial V}{\partial t}) W dV_0 = 0$$

This equation is difficult to solve for muscles in which the passive behavior could be described by an uncompressible Ogden model .

$W = \sum_{k=1}^{N} \frac{\mu_k}{\beta_k} \left( \tau_1^{\beta_k} + \tau_2^{\beta_k} + \tau_3^{\beta_k} - 3 \right)$ where the strain energy W is function of the three principal stretch ratios τ and other specific muscle's material parameters μ and β.

And the viscoelastic behavior is implemented by means of the Prony expansion, in such case the constitutive equation in the second Piola-Kirchhoff Stress tensor is:

$$S = (1 - \sum_{m=1}^{M} \delta_m) \frac{\partial W}{\partial E} + \sum_{m=1}^{M} \int_0^t \delta_m \frac{\partial W}{\partial E} e^{-(t-\varphi)/\theta_m} d\varphi$$

where E is the Green –Lagrange strain and $\delta_m$ and $\theta_m$ are viscoelastic muscle parameters.

From the next two figures ( indicative of the behavior of rigid bodies [ Saint Venant-Kirchhoff ] and soft bodies like muscles [Neo-Hookean ] ) it is possible to see that the behavior change abruptly with impact speed.

In fact at low speed there is no difference between the two collision, but increasing speed the Neo-Hookean material exhibits stiffening with increase in force and decreasing in time contact. In this simulation the parametric study of the time contact as function of the impact velocity shows that Neo-Hookean materials always yields lower impact time.

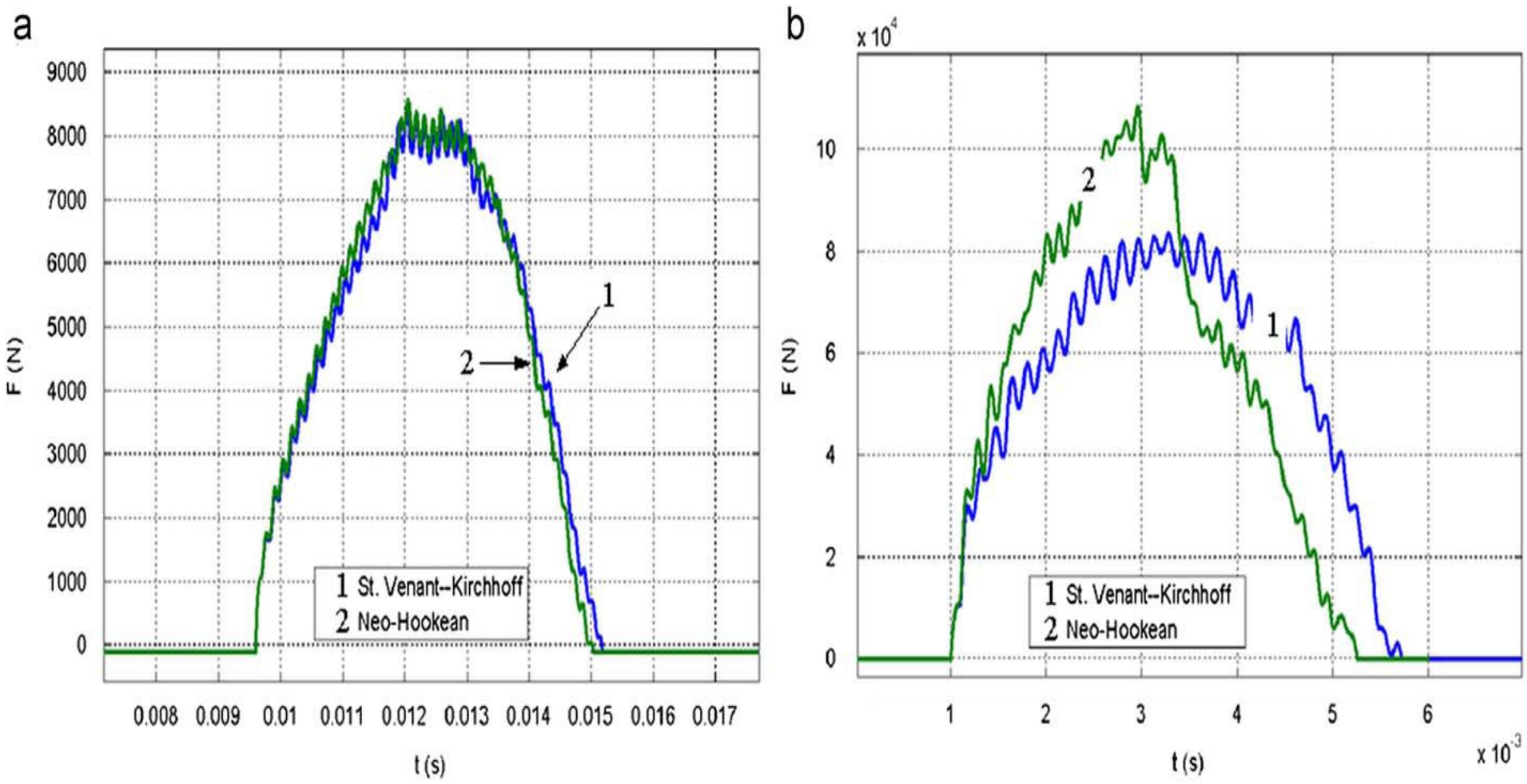

***Figg 21;22 Indicative different behavior of contact force at v=1m/s and v=10 m/s from [4] .***

## 8. *Interaction in Competition*

Interaction in Competition is basically performed studying both

a) Study of Interaction between athletes ( Throws ) application of knowledge on Kuzushi-Tsukuri Kake Phases

b) Study of the motion of “Couple of Athletes System”

The hard problem of interaction between fighters ( Remember that in Judo there are almost 80 different throwing techniques) was solved by the author and presented in Athens in 1987 grouping all judo techniques under two simple physical principles.
The study of the motion of “Couple of Athletes System” despite is “simple” locomotion aspect, is very complex one, and it was solved only by using a non linear approach, by the author, using the Matsumoto and co-workers. “Variograms” in the “Analysis of a Complex Physical Systems” ( 1990), in which it was demonstrated, for the first time, that the motion of “Couple of Athletes System” during fight belongs to the class of Brownian Motions.
In fact the whole system “Couple of Athletes” is surprisingly in stable equilibrium ( like a Brownian particle) and if the Couple is “still” in physics this system is defined “free” only external forces acting on it are the push/pull (random) forces applied by the adversary by means of friction under barefoot.
The push/pull forces, act for a very short time during the game’s time.
They influence only the quantity of motion variation ( like very short collision).
In formulas it is possible to write for each pull/push force:
$P = m\Delta v\, \delta\,(t-t')$
Then the general equation that manages the “Couple of Athletes System” motion in competition is the following Langevin Type of equation

$$m\dot{v} = -\mu v + m\Delta v(\pm 1)\delta(t-t')$$

Or in general form

$$F = -f + P$$

If the random Push/Pull force P have zero mean over time $\langle P \rangle = 0$

This means that push/pull forces can be considered as white noise, and motion belongs to the class of Brownian motions

In formula: $\frac{1}{2} m\overline{v}^2 = \eta \overline{O}^2$

Under this methodological approach, analysis of Judo fight sequences of still video frames, must be used in analysing both players’ aspect “complete movements” and “interactions”.
Interaction in judo throws is connected to the shifting speed of the Couple of Athletes and as already mentioned, founded on two main physical principles application:

1. ***Couple of forces application***

2. ***Physical lever application***

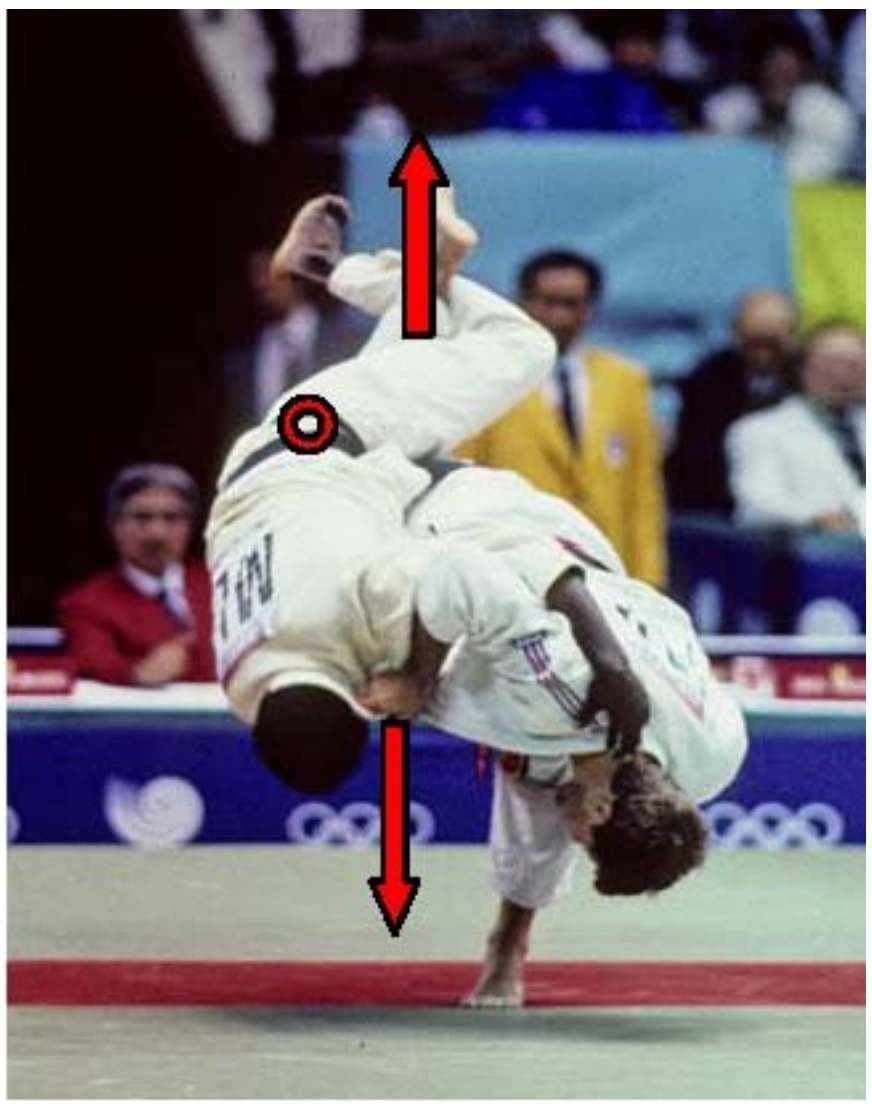

a)

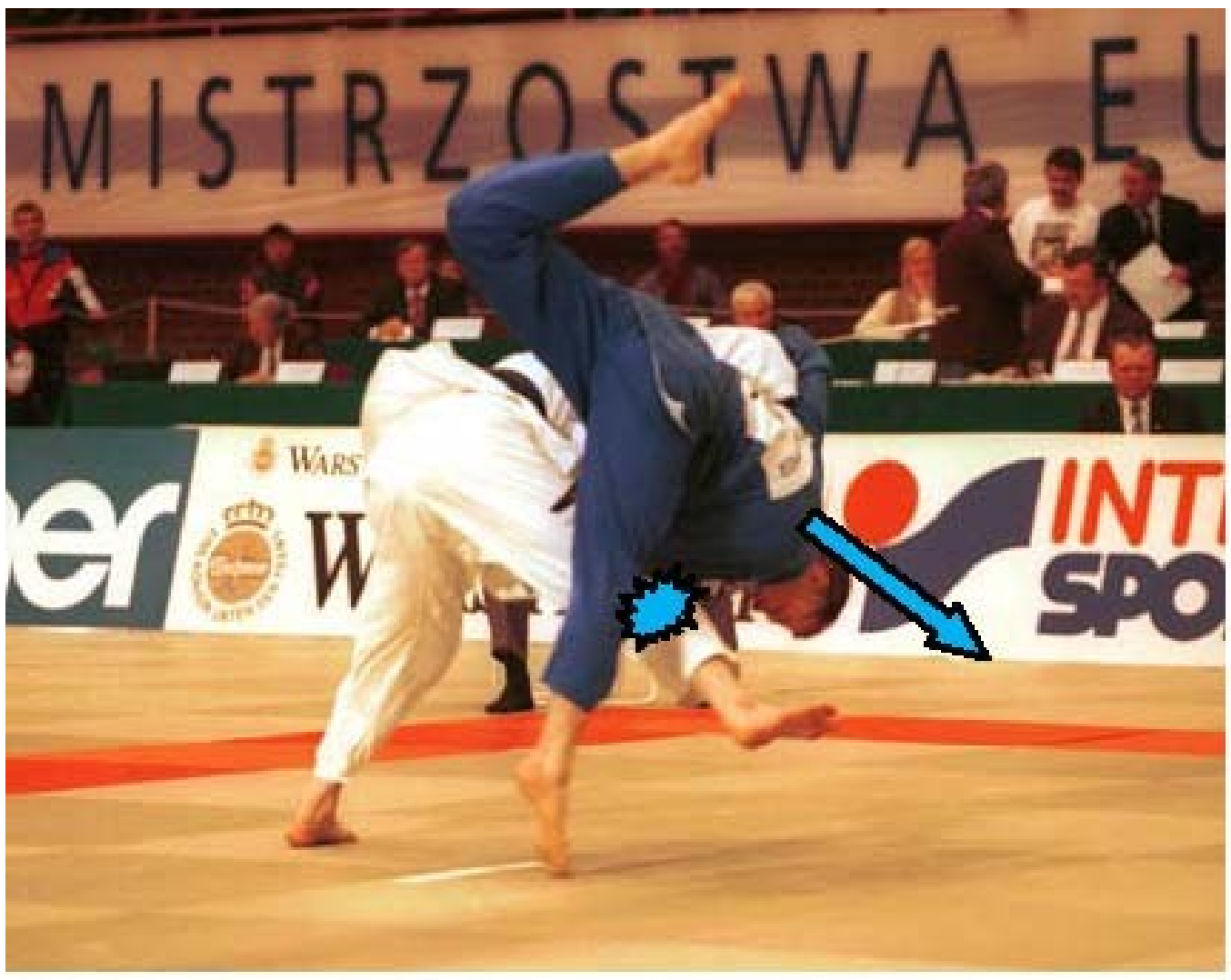

b)

***Figg.23,24 Throwing in competition,***
***a) Application of a couple of forces and b) Application of a physical lever***

Complete shifting movements that belong to the class of Brownian Motions are also connected to the today's most analyzed fighting phase *The Kuzushi Tzsukuri* movements in competition studied in the first part of the paper .

## *9. Conclusions*

In this paper we stated that Tsukuri Phase is condition necessary and sufficient for all kind of throws, that Kuzushi it is necessary not for all techniques, but only for the Physical lever group. We stated also that Tsukuri phases are composed by ***General Action Invariants ( GAI)*** equal for all, but for the physical lever group these ***General Action Invariants*** must be connected to some ***Specific Action Invariants (SAI)***, depending from both superior and inferior kinetic chains right positioning and motion.

As already remembered Kake phase, for the techniques of physical lever, is the result of the well interconnected work performed by both kinetic chains in different time steps: at first the superior chain open space for the body in the adversary grips, then the ***General Action*** (distance shortening) is developed followed harmonically by the connected work of both Inferior and ***Superior Action Invariants (SSAI)*** through the abdominals and trunk muscles.

These techniques need more skill and timing in harmonic chains connected movements than couple techniques.

In fact often they are ineffective, because one of the previous connected movements lack in harmony is able to keep Ippon.

These findings show us that the physical lever techniques are more complex as motor skill and, generally speaking, more expensive in energy terms.

Other important result is that couple of forces techniques does not depend necessarily by Kuzushi. Generally speaking, they can be applied only with Tsukuri movement without Kuzushi.

The formal independence from Kuzushi phase means that these techniques can be applied whatever is the Couple of Athletes shifting velocity.

In opposite the physical lever group needs to stop for a while the shifting velocity of couple of athletes to apply the Kuzushi for the techniques.

Then one important difference between Physical lever techniques and Couple techniques is that the first ones need of a special personal coordinative timing to the right application, in connection to the right positioning timing to apply an effective Kake phase the second ones need only a right positioning time, perhaps more complex than the first one, because the Couple of Athletes could be in motion.

Judo competition is a complex nonlinear system, and the study results show that all throws are grouped under two physical principles and the Couple of Athlete motion belongs to the class of Fractional Brownian Motions with well known aspecys of autosimilarity.